\documentstyle[epsfig,12pt]{article}
\catcode`\@=11
\long\def\@makefntext#1{
\protect\noindent \hbox to 3.2pt {\hskip-.9pt  
$^{{\ninerm\@thefnmark}}$\hfil}#1\hfill}		

\def\@makefnmark{\hbox to 0pt{$^{\@thefnmark}$\hss}}  
	
\def\ps@myheadings{\let\@mkboth\@gobbletwo
\def\@oddhead{\hbox{}
\rightmark\hfil\ninerm\thepage}   
\def\@oddfoot{}\def\@evenhead{\ninerm\thepage\hfil
\leftmark\hbox{}}\def\@evenfoot{}
\def\sectionmark##1{}\def\subsectionmark##1{}}

\renewcommand{\thefootnote}{\fnsymbol{footnote}}

\newcounter{sectionc}\newcounter{subsectionc}\newcounter{subsubsectionc}
\renewcommand{\section}[1] {\vspace*{0.6cm}\addtocounter{sectionc}{1} 
\setcounter{subsectionc}{0}\setcounter{subsubsectionc}{0}\noindent 
	{\normalsize\bf\thesectionc. #1}\par\vspace*{0.4cm}}
\renewcommand{\subsection}[1] {\vspace*{0.6cm}\addtocounter{subsectionc}{1} 
	\setcounter{subsubsectionc}{0}\noindent 
	{\normalsize\it\thesectionc.\thesubsectionc. #1}\par\vspace*{0.4cm}}
\renewcommand{\subsubsection}[1]
{\vspace*{0.6cm}\addtocounter{subsubsectionc}{1}
	\noindent {\normalsize\rm\thesectionc.\thesubsectionc.\thesubsubsectionc. 
	#1}\par\vspace*{0.4cm}}

\newcounter{appendixc}
\newcounter{subappendixc}[appendixc]
\newcounter{subsubappendixc}[subappendixc]

\renewcommand{\appendix}[1] {\vspace*{0.6cm}
        \refstepcounter{appendixc}
        \setcounter{figure}{0}
        \setcounter{table}{0}
        \setcounter{equation}{0}
        \renewcommand{\thefigure}{\Alph{appendixc}.\arabic{figure}}
        \renewcommand{\thetable}{\Alph{appendixc}.\arabic{table}}
        \renewcommand{\theappendixc}{\Alph{appendixc}}
        \renewcommand{\theequation}{\Alph{appendixc}.\arabic{equation}}
        \noindent{\bf Appendix \theappendixc #1}\par\vspace*{0.4cm}}

\def\abstracts#1{{
	\centering{\begin{minipage}{12.2truecm}\footnotesize\baselineskip=12pt\noindent
	\centerline{\footnotesize ABSTRACT}\vspace*{0.3cm}
	\parindent=0pt #1
	\end{minipage}}\par}} 

\renewenvironment{thebibliography}[1]
	{\begin{list}{\arabic{enumi}.}
	{\usecounter{enumi}\setlength{\parsep}{0pt}
\setlength{\leftmargin 1.25cm}{\rightmargin 0pt}
	 \setlength{\itemsep}{0pt} \settowidth
	{\labelwidth}{#1.}\sloppy}}{\end{list}}

\newcounter{itemlistc}
\newcounter{romanlistc}
\newcounter{alphlistc}
\newcounter{arabiclistc}

\newcommand{\fcaption}[1]{
        \refstepcounter{figure}
        \setbox\@tempboxa = \hbox{\footnotesize Fig.~\thefigure. #1}
        \ifdim \wd\@tempboxa > 6in
           {\begin{center}
        \parbox{6in}{\footnotesize\baselineskip=12pt Fig.~\thefigure. #1}
            \end{center}}
        \else
             {\begin{center}
             {\footnotesize Fig.~\thefigure. #1}
              \end{center}}
        \fi}

\newcommand{\tcaption}[1]{
        \refstepcounter{table}
        \setbox\@tempboxa = \hbox{\footnotesize Table~\thetable. #1}
        \ifdim \wd\@tempboxa > 6in
           {\begin{center}
        \parbox{6in}{\footnotesize\baselineskip=12pt Table~\thetable. #1}
            \end{center}}
        \else
             {\begin{center}
             {\footnotesize Table~\thetable. #1}
              \end{center}}
        \fi}

\def\@citex[#1]#2{\if@filesw\immediate\write\@auxout
	{\string\citation{#2}}\fi
\def\@citea{}\@cite{\@for\@citeb:=#2\do
	{\@citea\def\@citea{,}\@ifundefined
	{b@\@citeb}{{\bf ?}\@warning
	{Citation `\@citeb' on page \thepage \space undefined}}
	{\csname b@\@citeb\endcsname}}}{#1}}

\newif\if@cghi
\def\cite{\@cghitrue\@ifnextchar [{\@tempswatrue
	\@citex}{\@tempswafalse\@citex[]}}
\def\citelow{\@cghifalse\@ifnextchar [{\@tempswatrue
	\@citex}{\@tempswafalse\@citex[]}}
\def\@cite#1#2{{$\null^{#1}$\if@tempswa\typeout
	{IJCGA warning: optional citation argument 
	ignored: `#2'} \fi}}

 1
 1
 1

\font\ninerm=cmr9

\newcommand {\ignore}[1]{}
\def\lsim{\:\raisebox{-0.5ex}{$\stackrel{\textstyle<}{\sim}$}\:}
\def\gsim{\:\raisebox{-0.5ex}{$\stackrel{\textstyle>}{\sim}$}\:}
\newcommand{\Frac}[2]{\frac{\displaystyle #1}{\displaystyle #2}}
\def\thefootnote{\fnsymbol{footnote}}
\def\21{$SU(2) \ot U(1)$}
\def\321{$SU(3) \ot SU(2) \ot U(1)$}
\def\ne{\hbox{$\nu_e$ }}
\def\nm{\hbox{$\nu_\mu$ }}
\def\nt{\hbox{$\nu_\tau$ }} 
\def\ns{\hbox{$\nu_{s}$ }}

\def\apj#1#2#3{          { Astrophys. J. }{\bf #1}, #3 (19#2)}

\def\ib#1#2#3{           { ibid. }{\bf #1}, #3 (19#2)}

\def\pr#1#2#3{           { Phys. Rev. }{\bf #1}, #3 (19#2)}

\def\prl#1#2#3{          { Phys. Rev. Lett. }{\bf #1}, #3 (19#2)}

\def\n.c.#1#2#3{         { Nuovo Cim. }{\bf #1}, #3 (19#2)}
\def\r.n.c.#1#2#3{       { Riv. del Nuovo Cim. }{\bf #1}, #3 (19#2)}

\begin{document}
\hfill{\vbox{\hbox{IFT-P.049/98}
                \hbox{hep-ph/9808203}}}
\vskip 1cm
\centerline{\normalsize\bf Update on Atmospheric Neutrinos\cite{ourwork}\footnote{To Appear in Proceeding of the RINGBERG EUROCONFERENCE - 
NEW TRENDS IN NEUTRINO PHYSICS Ringberg Castle, Tegernsee, Germany, 
24 - 29 May 1998 }}
\baselineskip=25pt
\centerline{M.\ C.\ Gonzalez-Garcia}
\baselineskip=25pt
\centerline{\it Instituto de F\'{\i}sica Corpuscular - 
C.S.I.C/Universitat de Val\`encia. Spain}
\baselineskip=15pt
\centerline{and}
\centerline{\it Instituto de Fisica Teorica, 
Universidade Estadual Paulista}
\centerline{\it  Rua Pamplona 145,01405--900 S\~ao Paulo, Brazil}
\centerline{\footnotesize E-mail: concha.gonzalez@uv.es}
\vspace*{1cm}
\abstracts{I summarize here the results of a global 
fit to the full data set corresponding to $535$ days of
data of the Super-Kamiokande experiment as well as to all other
experiments in order to compare the two most likely solutions to the
atmospheric neutrino anomaly in terms of oscillations in the $\nu_\mu
\to \nu_\tau$ and $\nu_\mu \to \nu_s$ channels.} 

\normalsize\baselineskip=15pt
\setcounter{footnote}{0}
\renewcommand{\thefootnote}{\alph{footnote}}
\section{Introduction}

Atmospheric showers are initiated when primary cosmic rays hit the
Earth's atmosphere. Secondary mesons produced in this collision,
mostly pions and kaons, decay and give rise to electron and muon
neutrino and anti-neutrinos fluxes \cite{review}.  There has been a
long-standing anomaly between the predicted and observed $\nu_\mu$
$/\nu_e$ ratio of the atmospheric neutrino fluxes
\cite{atmexp}. Although the absolute individual $\nu_\mu$ or $\nu_e$
fluxes are only known to within $30\%$ accuracy, different authors
agree that the $\nu_\mu$ $/\nu_e$ ratio is accurate up to a $5\%$
precision. In this resides our confidence on the atmospheric neutrino
anomaly (ANA), now strengthened by the high statistics sample
collected at the Super-Kamiokande experiment \cite{superkam}.  
The most likely solution of the ANA involves neutrino
oscillations.  In principle we can invoke various neutrino oscillation
channels, involving the conversion of \nm into either \ne or \nt
(active-active transitions) or the oscillation of \nm into a sterile
neutrino \ns (active-sterile transitions). 
This last case is especially well-motivated theoretically, since
it constitutes one of the simplest ways to reconcile
\cite{4familytalks} the ANA with other puzzles in the
neutrino sector such as the solar neutrino problem as well
as the LSND result \cite{lsnd} and the possible need for a few eV mass
neutrino as the hot dark matter in the Universe \cite{SS1}.

The main aim of this talk is to compare the $\nu_\mu \to
\nu_\tau$ and the $\nu_\mu \to \nu_{s}$ transitions using the the 
new sample corresponding to 535 days of the Super-Kamiokande
data. This analysis uses the latest improved calculations of
the atmospheric neutrino fluxes as a function of zenith angle, 
including the muon polarization effect
and taking into account a variable neutrino production point
\cite{flux}.  

\section{Atmospheric Neutrino Oscillation Probabilities}

The expected neutrino event number both in the
absence and the presence of oscillations can be written as:
\begin{equation}
N_\mu= N_{\mu\mu} +\
 N_{e\mu} \; ,  \;\;\;\;\;\
N_e= N_{ee} +  N_{\mu e} \; ,
\label{eventsnumber}
\end{equation}
where
\begin{equation}
N_{\alpha\beta} = n_t T
\int
\frac{d^2\Phi_\alpha}{dE_\nu d(\cos\theta_\nu)} 
\kappa_\alpha(h,\cos\theta_\nu,E_\nu)
P_{\alpha\beta}
\frac{d\sigma}{dE_\beta}\varepsilon(E_\beta)
dE_\nu dE_\beta d(\cos\theta_\nu)dh\; .
\label{event0}
\end{equation}
and $P_{\alpha\beta}$ is the oscillation probability of $\nu_\beta \to
\nu_\alpha$ for given values of $E_{\nu}, \cos\theta_\nu$ and $h$,
i.e., $ P_{\alpha\beta} \equiv P(\nu_\alpha \to \nu_\beta; E_{\nu},
\cos\theta_\nu, h) $.  In the case of no oscillations, the
only non-zero elements are the diagonal ones,
i.e. $P_{\alpha\alpha}=1$ for all $\alpha$.

Here $n_t$ is the number of targets, $T$ is the experiment's running
time, $E_\nu$ is the neutrino energy and $\Phi_\alpha$ is the flux of
atmospheric neutrinos of type $\alpha=\mu ,e$; $E_\beta$ is the final
charged lepton energy and $\varepsilon(E_\beta)$ is the detection
efficiency for such charged lepton; $\sigma$ is the neutrino-nucleon
interaction cross section, and $\theta_\nu$ is
the angle between the vertical direction and the incoming neutrinos
($\cos\theta_\nu$=1 corresponds to the down-coming neutrinos).  In
Eq.~(\ref{event0}), $h$ is the slant distance from the production
point to the sea level for $\alpha$-type neutrinos with energy $E_\nu$
and zenith angle $\theta_\nu$. Finally, $\kappa_\alpha$ is the slant
distance distribution which is normalized to one \cite{flux}.

The neutrino fluxes, in particular in the sub-GeV range, depend on the
solar activity.  In order to take this fact into account in
Eq.~(\ref{event0}), a linear combination of atmospheric neutrino fluxes
$\Phi_\alpha^{max}$ and $\Phi_\alpha^{min}$, which correspond to the
most active Sun (solar maximum) and quiet Sun (solar minimum)
respectively, is used. 
  
For definiteness we assume a two-flavor oscillation scenario, in which
the $\nu_\mu$ oscillates into another flavour either $\nu_\mu \to
\nu_e$ , $\nu_\mu \to \nu_s$ or $\nu_\mu \to
\nu_\tau$. The Schr\"odinger evolution equation of the $\nu_\mu -\nu_X$ 
(where $X=e,\tau $ or $s$ sterile) system in the matter background for
{\sl neutrinos } is given by
\begin{eqnarray}
i{\mbox{d} \over \mbox{d}t}\left(\matrix{
\nu_\mu \cr\ \nu_X\cr }\right) & = & 
 \left(\matrix{
 {H}_{\mu}
& {H}_{\mu X} \cr
 {H}_{\mu X} 
& {H}_X \cr}
\right)
\left(\matrix{
\nu_\mu \cr\ \nu_X \cr}\right) \,\,, \\
  H_\mu & \! = &  \! 
 V_\mu + \frac{\Delta m^2}{4E_\nu} \cos2 \theta_{\mu X}\,, \,\,\,\,\,\,\,\,\,
H_X \!= V_X -  \frac{\Delta m^2}{4E_\nu} \cos2 \theta_{\mu X}, \nonumber \\
H_{\mu X}& \!= &  - \frac{\Delta m^2}{4E_\nu} \sin2 \theta_{\mu X} \nonumber
\label{evolution1}
\end{eqnarray}
where 
\begin{eqnarray}
\label{potential}
V_\tau=V_\mu = \frac{\sqrt{2}G_F \rho}{M} (-\frac{1}{2}Y_n)\,, 
& \;\;\;\;\; & V_s=0 \nonumber \\
V_e = \frac{\sqrt{2}G_F \rho}{M} ( Y_e - \frac{1}{2}Y_n) & & 
\nonumber
\end{eqnarray}
Here $G_F$ is the Fermi constant, $\rho$ is the matter density at the
Earth, $M$ is the nucleon mass, and $Y_e$ ($Y_n$) is the electron
(neutron) fraction. We define $\Delta m^2=m_2^2-m_1^2$ in such a way
that if $\Delta m^2>0 \: (\Delta m^2<0)$ the neutrino with largest
muon-like component is heavier (lighter) than the one with largest
X-like component. For anti-neutrinos the signs of potentials $V_X$
should be reversed. We have used the approximate analytic expression
for the matter density profile in the Earth obtained in
ref. \cite{lisi}. In order to obtain the oscillation probabilities
$P_{\alpha\beta}$ we have made a numerical integration of the
evolution equation. The probabilities for neutrinos and anti-neutrinos
are different because the reversal of sign of matter potential. Notice
that for the $\nu_\mu\to\nu_\tau$ case there is no matter effect while
for the $\nu_\mu\to\nu_s$ case we have two possibilities depending on
the sign of $\Delta m^2$.  For $\Delta m^2 > 0$ the matter efects
enhance {\sl neutrino} oscillations while depress {\sl antineutrino}
oscillations, whereas for the other sign ($\Delta m^2<0$) the opposite
holds. The same occurs also for $\nu_\mu\to\nu_e$.  Although in the
latter case one can also have two possible signs, we have chosen the
most usually assumed case where the muon neutrino is heavier than the
electron neutrino, as it is theoretically more appealing. Notice also
that, as seen later, the allowed region for this sign is larger than
for the opposite, giving the most conservative scenario when comparing
with the present limits from CHOOZ.

\section{Atmospheric Neutrino Data Fits} 

Here I describe our fit method to determine the atmospheric
oscillation parameters for the various possible oscillation channels,
including matter effects for both $\nu_\mu \to \nu_e$ and $\nu_\mu \to
\nu_s$ channels. 
The steps required in order to generate the allowed regions of
oscillation parameters were given in ref. \cite{ourwork}. 
I will comment only that when combining the results of the experiments we
do not make use of the double ratio, $R_{\mu/e}/R^{MC}_{\mu/e}$, but
instead we treat the $e$ and $\mu$-like data separately, taking into
account carefully the correlation of errors. It is well-known that the
double ratio  is  not well suited from a statistical point of view
due to its non-Gaussian character. Thus,
following ref. \cite{ourwork,fogli2} we define the $\chi^2$ as
\begin{equation}
\chi^2 \equiv \sum_{I,J}
(N_I^{data}-N_I^{theory}) \cdot 
(\sigma_{data}^2 + \sigma_{theory}^2 )_{IJ}^{-1}\cdot 
(N_J^{data}-N_J^{theory}),
\label{chi2}
\end{equation}
where $I$ and $J$ stand for any combination of the experimental data
set and event-type considered, i.e, $I = (A, \alpha)$ and $J = (B,
\beta)$ where, $A,B$ stands for Fr\'ejus, Kamiokande sub-GeV, IMB,... and
$\alpha, \beta = e,\mu$.  In Eq.~(\ref{chi2}) $N_I^{theory}$ is the
predicted number of events calculated from Eq.~(\ref{eventsnumber})
whereas $N_I^{data}$ is the number of observed events.  In
Eq.~(\ref{chi2}) $\sigma_{data}^2$ and $\sigma_{theory}^2$ are the
error matrices containing the experimental and theoretical errors
respectively. They can be written as
\begin{equation}
\sigma_{IJ}^2 \equiv \sigma_\alpha(A)\, \rho_{\alpha \beta} (A,B)\,
\sigma_\beta(B),
\end{equation}
where $\rho_{\alpha \beta} (A,B)$ stands for the correlation between
the $\alpha$-like events in the $A$-type experiment and $\beta$-like
events in $B$-type experiment, whereas $\sigma_\alpha(A)$ and
$\sigma_\beta(B)$ are the errors for the number of $\alpha$ and
$\beta$-like events in $A$ and $B$ experiments, respectively.

We compute $\rho_{\alpha \beta} (A,B)$ as in ref. \cite{fogli2}.  A
detailed discussion of the errors and correlations used in the
analysis can be found in Ref.\cite{ourwork}.  
We have conservatively ascribed a 30\% uncertainty to the
absolute neutrino flux, in order to generously account for the spread
of predictions in different neutrino flux calculations.  Next we
minimize the $\chi^2$ function in Eq.~(\ref{chi2}) and determine the
allowed region in the $\sin^22\theta-\Delta m^2$ plane, for a given
confidence level, defined as,
\begin{equation}
\chi^2 \equiv \chi_{min}^2  + 4.61\ (9.21)\   \  
\  \mbox{for}\  \ 90\  (99) \% \ \  \mbox{C.L.}
\label{chimin}
\end{equation}

\begin{figure}
\centerline{\protect\hbox{\epsfig{file=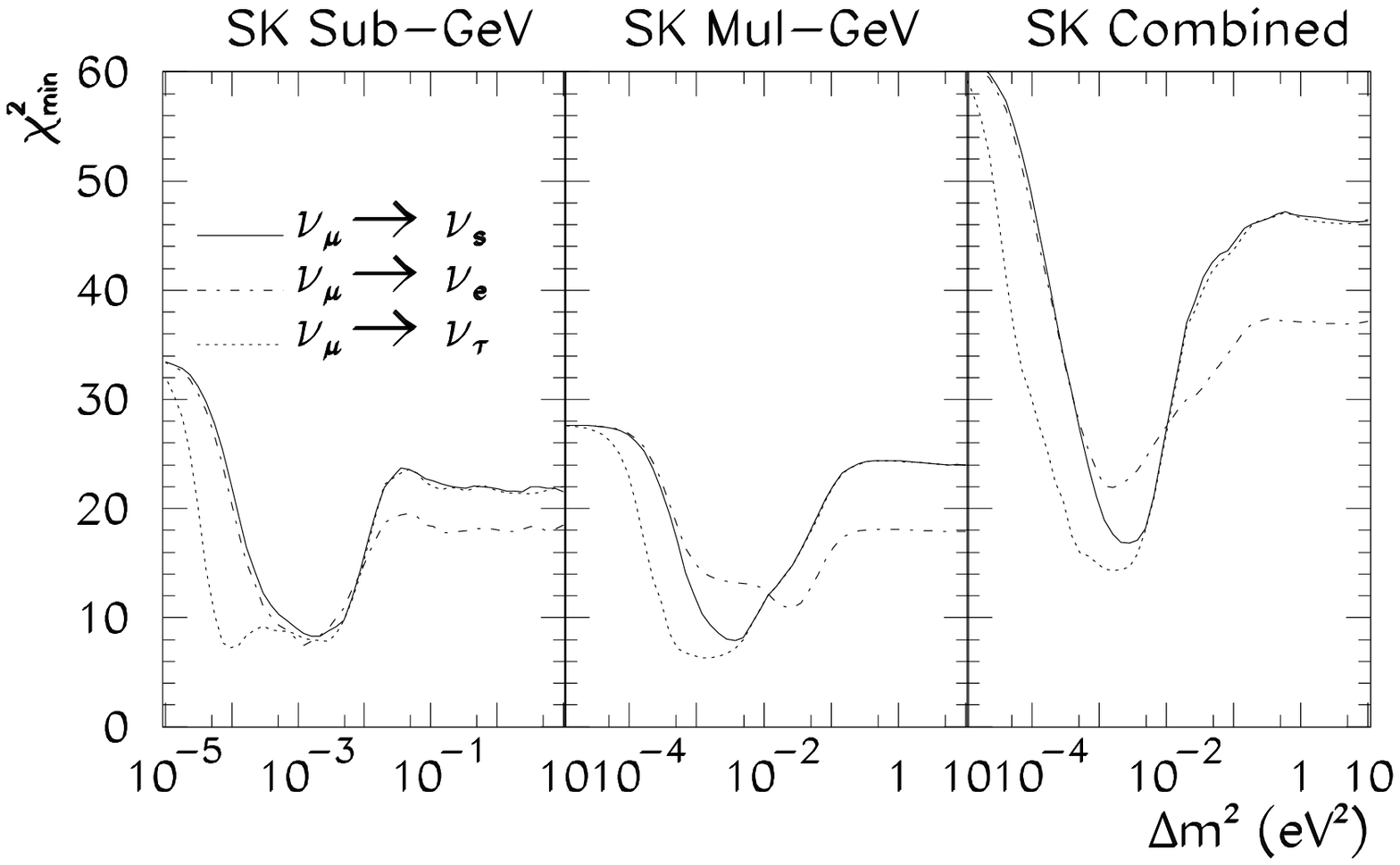,width=0.7\textwidth,height=0.3\textheight}}}
\fcaption{$\chi^2_{min}$ for fixed $\Delta m^2$ versus $\Delta m^2$ for
each oscillation channel for Super-Kamiokande sub-GeV and multi-GeV data, 
and for the combined sample. Since the minimum is always obtained
close to maximum mixing the curves for $\nu_\mu\rightarrow \nu_s$ for both
signs of $\Delta m^2$ coincide.}
\label{chimin1} 
\end{figure} 
In Fig.~\ref{chimin1} we plot the minimum $\chi^2$ (minimized with respect
to  $\sin^2 2\theta$) as a function of $\Delta m^2$.  Notice that for large 
$\Delta m^2 \gsim 0.1$ eV$^2$, the $\chi^2$ is nearly constant.  This happens
because in this limit the contribution of the matter potential in
Eq~(\ref{evolution1}) can be neglected with respect to the $\Delta
m^2$ term, so that the matter effect disappears and moreover, the
oscillation effect is averaged out. In fact one can see that in this
range we obtain nearly the same $\chi^2$ for the $\nu_\mu \to
\nu_\tau$ and $\nu_\mu
\to \nu_{s}$ cases. For very small $\Delta m^2 \lsim 10^{-4}$ eV$^2$,
the situation is opposite, namely the matter term dominates and we
obtain a better fit for the $\nu_\mu \to \nu_\tau$ channel, as can be
seen by comparing the $\nu_\mu \to \nu_{\tau}$ curve of the
Super-Kamiokande sub-GeV data (dotted curve in the left panel of
Fig.~\ref{chimin1}) with the $\nu_\mu \to \nu_s$ and 
$\nu_\mu \to \nu_e$ curves in the left panel of
Fig.~\ref{chimin1}). For extremely small $\Delta m^2 \lsim 10^{-4}$
eV$^2$, values $\chi^2$ is quite large and approaches a constant,
independent of oscillation channel, as in the no-oscillation case.
Since the average energy of Super-Kamiokande multi-GeV data is higher
than the sub-GeV one, we find that the limiting $\Delta m^2$ value
below which $\chi^2$ approaches a constant is higher, as seen in the
middle panel. Finally, the right panel in Fig.~\ref{chimin1} is
obtained by combining sub and multi-GeV data.
\begin{table}[h]
\tcaption{Minimum value of $\chi^2$ and the best fit point for each 
oscillation channel and for different data sets. 
For
$\nu_\mu\rightarrow \nu_s$ the minimum $\chi^2$ is practically independent
of the sign of $\Delta m^2$ as the minimum is located at maximum mixing 
angle.}
\label{tab:data}
\begin{center}
\begin{tabular}{|l|l|l|l|l|l|}
\hline
Experiment &  & $\nu_{\mu} \to  \nu_\tau$ &  
                $\nu_{\mu} \to  \nu_s$   &
                $\nu_{\mu} \to  \nu_e$  
\\\hline

Super-Kam   & $\chi^2_{min}$ 
                           & $  7.1 $ & $  8.2 $ & $ 7.3$
\\
 sub-GeV                          & $ \Delta m^2 $ ( $10^{-3} $eV$^2$ )
                           & $ 0.11 $    & $1.9$   & $1.2$
\\
                           & $\sin^2 2\theta$ 
                           &  $1.0$   & $1.0$   & $0.97$ 
\\\hline

Super-Kam  & $\chi^2_{min}$ 
                           & $  6.3$ & $  7.9 $ & $ 10.8$
\\
  multi-GeV                 & $ \Delta m^2 $ ( $10^{-3} $eV$^2$ )
                           & $1.5$    & $3.5$   & $24.7$
\\
                           & $\sin^2 2\theta$ 
                           &  $0.97$   & $1.0$   & $0.72$
                           \\\hline 
Super-Kam   & $\chi^2_{min}$ 
                           &$ 14.3$ & $ 16.8 $ & $21.8$
\\
    Combined              & $ \Delta m^2 $ ( $10^{-3} $eV$^2$ ) 
                           & $1.6$    & $2.6$   & $1.5$
\\
                           & $\sin^2 2\theta$ 
                           &  $1.0$   & $1.0$   & $0.97$ 
\\\hline
All experiments    & $\chi^2_{min}$ 
                           &$ 47.2$ & $ 48.6 $ & $48.6$
\\
  Combined               & $ \Delta m^2 $ ( $10^{-3} $eV$^2$ ) 
                           & $2.9$    & $3.5$  & $3.0$
\\
                           & $\sin^2 2\theta$ 
                           &  $1.0$   & $1.0$   & $0.99$ 
\\\hline
\end{tabular}
\end{center}
\end{table}
A last point worth commenting is that for the $\nu_\mu \to \nu_{\tau}$
case in the sub-GeV sample there are two almost degenerate values of
$\Delta m^2$ for which $\chi^2$ attains a minimum while for the multi-GeV 
case there is just one minimum at $1.5 \times 10^{-3} $eV$^2$. Finally
in the third panel in Fig.~\ref{chimin1} we can see that by
combining the Super-Kamiokande sub-GeV and multi-GeV data we have a
unique minimum at $1.6 \times 10^{-3} $eV$^2$.

\section{Results for the Oscillation Parameters}
 
\begin{figure}
\centerline{\protect\hbox{\epsfig{file=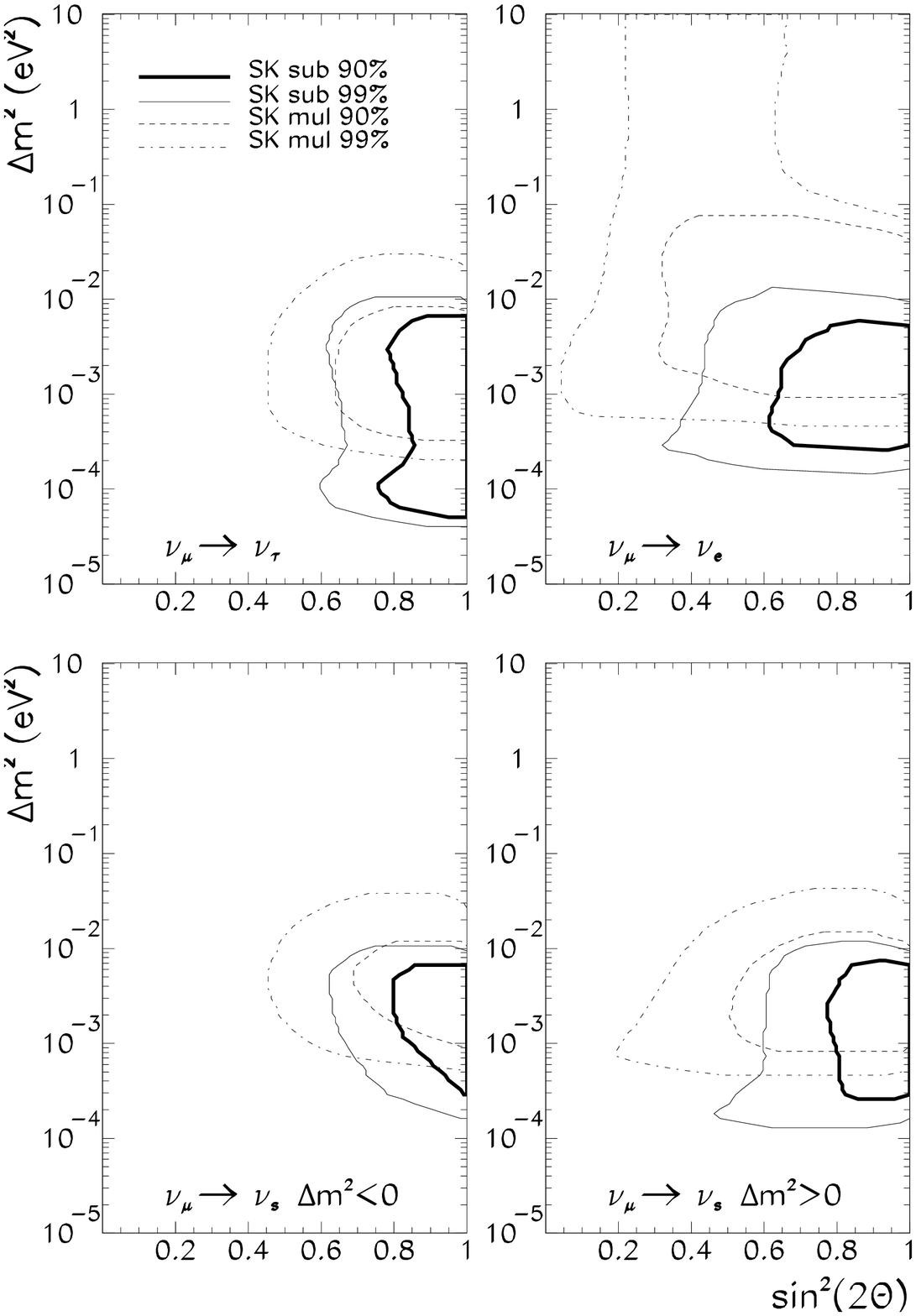,width=0.6\textwidth,height=0.4\textheight}}}
\fcaption{
Allowed regions of oscillation parameters for Super-Kamiokande for
the different oscillation channels as labeled in the figure.
In each panel, we show the allowed regions for the sub-GeV data
at 90 (thick solid line) and 99 \% CL  (thin solid line)
and the multi-GeV data at 90 (dashed line) and 99 \% CL  (dot-dashed line).}
\label{mutausk1}
\end{figure}

The results of our $\chi^2$ fit of the Super-Kamiokande sub-GeV and
multi-GeV atmospheric neutrino data are given in Fig.~\ref{mutausk1}.
In this figure we give the allowed region of oscillation parameters at 90
and 99 \% CL.  
One can notice that the matter effects are similar for the upper right
and lower right panels because matter effects enhance the oscillations
for {\sl neutrinos} in both cases.  In contrast, in the case of
$\nu_\mu \to \nu_s$ with $\Delta m^2<0$ the enhancement occurs only
for {\sl anti-neutrinos} while in this case the effect of matter
suppresses the conversion in $\nu_\mu$'s.  Since the yield of
atmospheric neutrinos is bigger than that of 
anti-neutrinos, clearly the matter effect suppresses the overall
conversion probability. Therefore we need in this case a larger value
of the vacuum mixing angle, as can be seen by comparing the left and
right lower panels in Fig.~\ref{mutausk1}.

Notice that in all channels where matter effects play a role 
the range of acceptable $\Delta m^2$ is
shifted towards larger values, when compared with the $\nu_\mu \to
\nu_\tau$ case. This follows from looking at the relation between 
mixing {\sl in vacuo} and in matter. In fact, away from the
resonance region, independently of the sign of the matter potential,
there is a suppression of the mixing inside the Earth. As a result,
there is a lower cut in the allowed $\Delta m^2$ value, and it lies
higher than what is obtained in the data fit for the $\nu_\mu \to
\nu_\tau$ channel.  

It is also interesting to analyse the effect of combining the
Super-Kamiokande sub-GeV and multi-GeV atmospheric neutrino data.
Comparing the results obtained with 535 days given in  
the table above with those obtained with 325 days 
of Super-Kamiokande\cite{ourwork} we see that the allowed region is
relatively stable with respect to the increased 
statistics.  However, in contrast to the case for 325.8 days, now the
$\nu_{\mu} \to \nu_\tau$ channel is as good as the $\nu_{\mu} \to
\nu_e$, when only the sub-GeV sample is included, with a clear 
Super-Kamiokande preference for the $\nu_{\mu} \to \nu_\tau$
channel. As before, the combined sub-GeV and multi-GeV data prefers
the $\nu_{\mu} \to \nu_X$, where $X=\tau$ or {\sl sterile}, over the
$\nu_{\mu} \to \nu_e$ solution.

\begin{figure}
\centerline{\protect\hbox{\epsfig{file=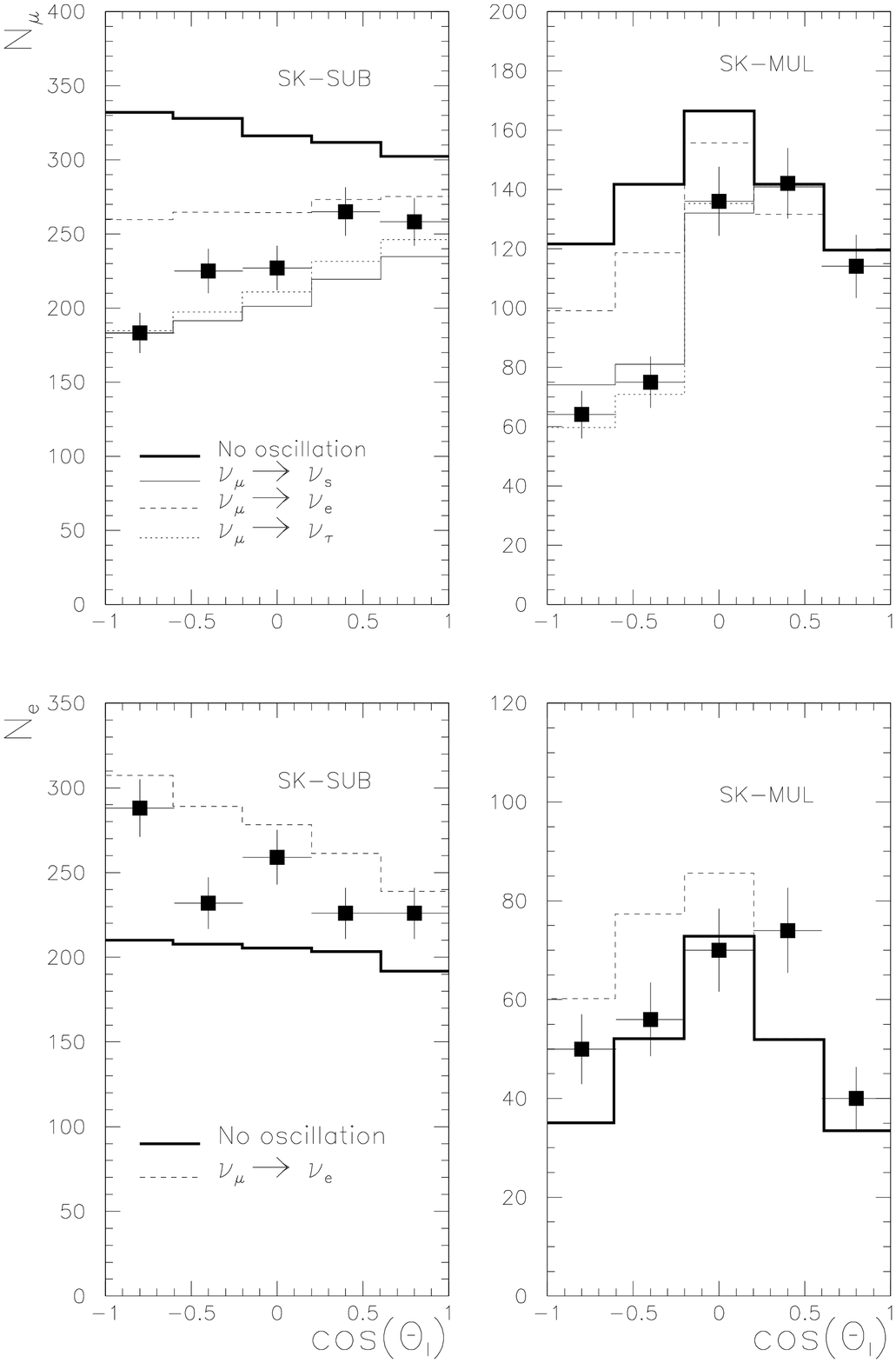,width=0.5\textwidth,height=0.4\textheight}}}
\caption{Angular distribution for Super-Kamiokande electron-like and muon-
like sub-GeV and multi-GeV events together with our prediction in the
absence of oscillation (dot-dashed) as well as the prediction for the
best fit point for $\nu_\mu \to \nu_s$ (solid line), $\nu_\mu \to
\nu_e$ (dashed line) and $\nu_\mu \to \nu_\tau$ (dotted line)
channels.  The error displayed in the experimental points is only
statistical.}
\label{ang_mu}  
\end{figure}

To conclude this section I now turn to the predicted zenith angle distributions
for the various oscillation channels. As an example we take the case
of the Super-Kamiokande experiment and compare separately the sub-GeV
and multi-GeV data with what is predicted in the case of
no-oscillation (thick solid histogram) and in all oscillation channels
for the corresponding best fit points obtained for the {\sl combined}
sub and multi-GeV data analysis performed above (all other
histograms). This is shown in Fig.~\ref{ang_mu}.  

It is worthwhile to see why the $\nu_\mu \to \nu_e$ channel is bad for
the Super-Kamiokande multi-GeV data by looking at the upper right
panel in Fig.~\ref{ang_mu}. Clearly the zenith distribution predicted
in the no oscillation case is symmetrical in the zenith angle very
much in disagreement with the data. In the presence of $\nu_\mu \to
\nu_e$ oscillations the asymmetry in the distribution is much smaller than
in the $\nu_\mu \to \nu_\tau$ or $\nu_\mu \to \nu_s$ channels, as seen
from the figure.
Also since the best fit point for $\nu_\mu \to
\nu_s$ occurs at $\sin(2\theta)=1$, the corresponding distributions
are independent of the sign of $\Delta m^2$.  

\section{Atmospheric versus Accelerator and Reactor Experiments}

\begin{figure}
\centerline{\protect\hbox{\epsfig{file=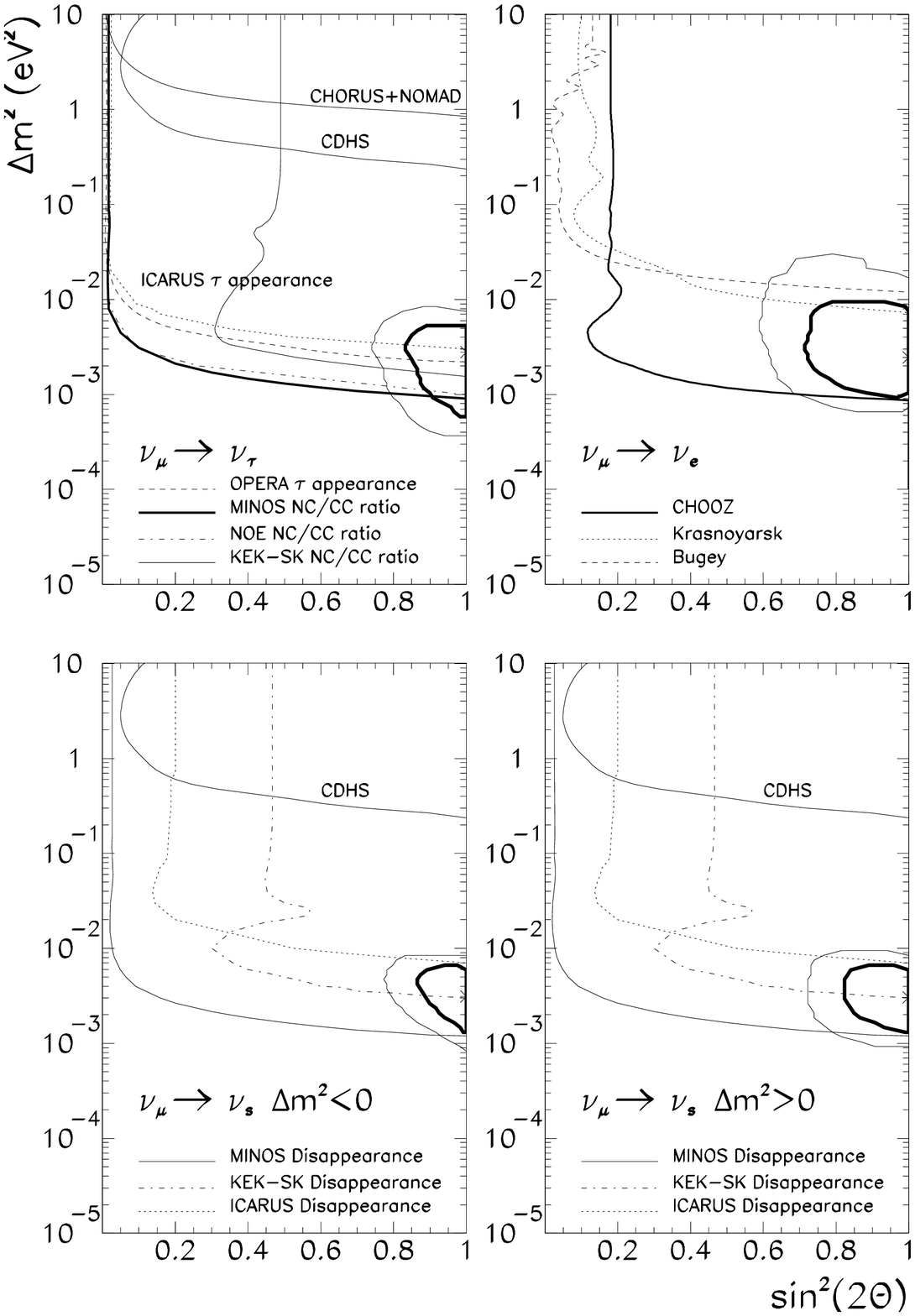,width=0.8\textwidth,height=0.8\textheight}}}
\fcaption{
Allowed oscillation parameters for all experiments combined at 90
(thick solid line) and 99 \% CL (thin solid line) for each oscillation
channel as labeled in the figure.  We also display the expected
sensitivity of the present accelerator and reactor experiments as well
as to future long-baseline experiments in each channel.
The best fit point is marked with a star.}
\label{mutausk4} 
\end{figure}

I now turn to the comparison of the information obtained from the
analysis of the atmospheric neutrino data presented above with the
results from reactor and accelerator experiments as well as the
sensitivities of future experiments. For this purpose I present the
results obtained by combining all the experimental atmospheric
neutrino data from various experiments\cite{atmexp}. In
Fig.~\ref{mutausk4} we show the combined information obtained from
our analysis of all atmospheric neutrino data involving
vertex-contained events and compare it with the constraints from
reactor experiments such as Krasnoyarsk, Bugey, and CHOOZ\cite{reactors}, 
and the accelerator experiments such as CDHSW, CHORUS, and NOMAD
\cite{accelerators}. We also include in the same figure the sensitivities
that should be attained at the future long-baseline experiments now
under discussion.

The first important point is that from the upper-right panel of
Fig.~\ref{mutausk4} one sees that the CHOOZ reactor\cite{reactors} data
already exclude completely the allowed region for the $\nu_{\mu} \to
\nu_e$ channel when  all experiments are  combined at 90\% CL. The situation
is different if only the combined sub-GeV and multi-GeV
Super-Kamiokande are included. In such a case the region obtained
is not completely excluded
by CHOOZ at 90\% CL.  Present accelerator experiments are not very
sensitive to low $\Delta m^2$ due to their short baseline. As a
result, for all channels other than $\nu_{\mu} \to \nu_e$ the present
limits on neutrino oscillation parameters from CDHSW,
CHORUS and NOMAD \cite{accelerators} are fully consistent with the region
indicated by the atmospheric neutrino analysis. Future long baseline
(LBL) experiments have been advocated as a way to independently check
the ANA.  Using different tests such long-baseline experiments now
planned at KEK (K2K) \cite{chiaki}, Fermilab (MINOS) \cite{minos} and
CERN ( ICARUS \cite{icarus}, NOE \cite{noe} and OPERA \cite{opera}) would 
test the pattern
of neutrino oscillations well beyond the reach of present
experiments. 
These tests are the following: $\tau$ appearance searches, 
$NC/CC$ ratio which measures $\Frac{(NC/CC)_{near}}{(NC/CC)_{far}}$, and 
the muon disappearance or $CC_{near}/CC_{far}$ test.
The second test can potentially discriminate between the active and sterile
channels, i.e. $\nu_\mu \to \nu_\tau$ and $\nu_\mu \to \nu_s$. However
it cannot discriminate between $\nu_\mu \to \nu_s$ and the
no-oscillation hypothesis.  In contrast, the last test can probe the
oscillation hypothesis itself. Notice that the sensitivity curves 
corresponding to the disappearance
test labelled as {\sl KEK-SK Disappearance } at the lower panels of
Fig.~\ref{mutausk4} are the same for the $\nu_\mu \to \nu_\tau$ and
the sterile channel since the average energy of KEK-SK is too low to
produce a tau-lepton in the far detector. 
In contrast the MINOS experiment has a higher
average initial neutrino energy and it can see the tau's. Although in
this case the exclusion curves corresponding to the disappearance test
are in principle different for the different oscillation channels, in
practice, however, the sensitivity plot is dominated by the systematic
error.  As a result discriminating between $\nu_\mu \to
\nu_{\tau}$ and $\nu_\mu \to \nu_s$ would be unlikely with the
Disappearance test.

In summary we find that the regions of oscillation
parameters obtained from the analysis of the atmospheric neutrino data
on vertex-contained events cannot be fully tested by the LBL
experiments, when the Super-Kamiokande data are included in the fit for
the $\nu_\mu \to
\nu_\tau$ channel as can be seen clearly from the upper-left panel
of Fig.~\ref{mutausk4}. 
One might expect that, due to the upward shift
of the $\Delta m^2$ indicated by the fit for the sterile case, it
would be possible to completely cover the corresponding region of
oscillation parameters. This is the case for the MINOS 
disappearance test. But in general since only the disappearance 
test can discriminate against the no-oscillation hypothesis, and this
test is intrinsically weaker due to systematics, we find
that also for the sterile case most of the LBL experiments can not completely
probe the region of oscillation parameters indicated by the
atmospheric neutrino analysis. This is so irrespective of the sign of
$\Delta m^2$: the lower-left panel in Fig.~\ref{mutausk4} shows the
$\nu_\mu \to \nu_s$ channel with $\Delta m^2<0$ while the $\nu_\mu \to
\nu_s$ case with $\Delta m^2>0$ is shown in the lower-right panel.

I am very grateful to the Instituto de F\'{\i}sica Te\'orica of the 
Universidade Estadual Paulista where these proceedings were written 
for its kind hospitality during my visit. 
This work was supported by DGICYT under
grant PB95-1077, by CICYT under  grant AEN96-1718, and by Funda\c{c}\~ao
de Amparo \`a Pesquisa do Estado de S\~ao Paulo.

\end{document}